\documentclass[hyper]{JHEP} 

\usepackage{epsfig}

\newcommand\fverb{\setbox\pippobox=\hbox\bgroup\verb}
\newcommand\fverbdo{\egroup\medskip\noindent%
			\fbox{\unhbox\pippobox}\ }
\newcommand\fverbit{\egroup\item[\fbox{\unhbox\pippobox}]}
\newbox\pippobox

\title{D-Branes from N Non-BPS D0-Branes}

\author{by J. Kluso\v{n}\\
	 Department of Theoretical Physics and Astrophysics\\
                   Faculty of Science, Masaryk University\\
Kotl\'{a}\v{r}sk\'{a} 2, 611 37, Brno\\
Czech Republic\\
	E-mail: \email{klu@physics.muni.cz}}

\preprint{\hepth{0009189}}

\abstract{In this paper we would like to show that from
$N$ non-BPS D0-branes in Type IIB theory
 we can obtain all BPS and non-BPS 
D-branes through tachyon condensation in
the limit $N\rightarrow \infty$.}


\def\tr{\mathrm{Tr}}
\def\str{\mathrm{Str}}

\def\I{\mathbf{i}}

\def\bra #1{\left<#1\right|}
\def\ket #1{\left|#1\right>}
\begin{document}
\section{Introduction}\label{first}
In the recent years there was a significant progress in
the understanding of the unstable configurations in 
superstring theories. This work has been pioneered with
the seminar papers by A. Sen \cite{SenP}. It was
argued in \cite{witen,Horava} that all D-branes can
arise as solitonic solutions in the world-volume theory of
the unstable configurations of D-branes. (For review of the
relation between K-theory and D-branes, see 
\cite{Olsen}, for recent discussion, see \cite{witen2,Matsuo,
Moore}.)

Evidence for this proposal was given from the
analysis of CFT description of this system \cite{SenP},
for review of this approach, see \cite{Lerda,
Schwarz}. It was also shown on many examples that
string field theory approach to this problem is very
effective one which allows to calculate tachyon potential
\cite{SenFT1,SenFT2,BerkovitzFT1,HarveyFT,SenFT3,
TaylorFT,KochFT,Desmet,NaqviFT,SenFT4,DavidFT,WittenFT,
RasteliFT,SenFT5,TaylorFT2,KochFT2,NaqviFT2}. Recently
this problem was also studied from the point of view 
of the Witten's background independent open string
field theory \cite{WittenBT,WittenBT1,WittenBT2,ShatasviliBT,
ShatasviliBT1,ShatasviliBT2,MooreBT,SenBT}.
Success of string field theory in the analysis of 
tachyon condensation indicates that string field theory could play
more fundamental role in the nonperturbative formulation of string theory.

The second approach to the problem of tachyon condensation
is based on the noncommutative geometry \cite{WittenNG}.
This analysis has been inspired with the seminal paper
\cite{StromingerA}. Application of  this approach to the problem
of the tachyon condensation was pioneered in
\cite{Gopakumar,Harvey}. This research was then developed
in other papers \cite{Harvey2,Strominger2,Rey,Matsuo,
SenTP,Mukhi}.

In this paper we would like to discuss the problem of
the tachyon condensation from slightly different point of view.
We would like to show that nontrivial  tachyon condensation is
also possible in the action for $N$ non-BPS D0-branes, which
results in the emergence of higher dimensional BPS and 
non-BPS D-branes in the process similar to the emergence
of higher dimensional branes in Matrix theory \cite{BanksM,
Banks} (For review, see \cite{Taylor,BanksR1,BanksR2,Taylor2}.).
However, there is an important difference. In matrix theory 
we  work with the exact form of the action and with 
the maximal supersymmetric theory which allows to obtain exact result.
On the other hand, we do not know exactly the form
of a non-BPS D-brane action  which can be guessed only on some general grounds.
Possible form of this action was proposed in seminal paper
\cite{SenNA}, other attempts to define this action  appeared in \cite{KlusonP,KlusonP1,KlusonP2,
Garousi,Bergshoff,Kluson}. We also study system with
maximally broken supersymmetry so that the analysis is much more
difficult.  However, we still believe that our
approach is  useful since it presents an evidence
of the emergence of higher dimensional D-branes from
lower dimensional ones thanks to the tachyon 
condensation.  As we will see the resulting configurations carry
the correct RR charges which allows us to expect that these
solutions are well defined.

We start in the section (\ref{second}) with discussion of the
 bosonic form of the action for $N$ non-BPS D9-branes.
Then we use T-duality transformation, following \cite{Garousi},
in order to obtain an action for $N$ non-BPS D0-branes.

In section (\ref{third}) we will discuss possible applications of
this action. Since we 
do not know the exact form of this action, we restrict ourselves
only to the linear approximation.  We also show that the whole action
(without restriction to the linear approximation)
contains the solution corresponding to collection of non-BPS
D0-branes. We also show that this action has a solution
corresponding to the tensionless D0-branes, discussed recently in
 the case of noncommutative field theories \cite{Gopakumar,
Harvey}. 

In sections (\ref{fourth}) and (\ref{fifth}) we will show
that from the collection of $N$ non-BPS D-branes
we can obtain all BPS and non-BPS D-branes in the limit $N\rightarrow
\infty$ when we can replace infinite dimensional matrices
with operators acting on some abstract Hilbert space. We will
proceed in the same way as in the study of tachyon condensation
in noncommutative theories \cite{Gopakumar,Harvey,Strominger2}
and we will show that these D-branes carry nonzero RR-charge 
thanks to the existence of generalised Wess-Zumino term
\cite{Mayers}. 

In the next section we 
start with the action for non-BPS D0-branes, which can
be obtained from the action for non-BPS D9-branes
through  T-duality transformation.
\section{Non-BPS D-brane action}\label{second}

In this section we will discuss the possible
form of the action for $N$ non-BPS D-branes,
following the seminal paper 
\cite{SenNA}. Similar discussions 
were presented in \cite{KlusonP,KlusonP1,
KlusonP2,Garousi,Bergshoff}.
 
We start with the most general form
\footnote{We mean the most general form
up to the first derivatives. Of course, there could
be infinite number of higher derivatives. In the case
when commutators are small the action with
the first derivatives is the appropriate one.} of the
action for $N$ non-BPS D9-branes in the form
\begin{equation}\label{Action}
S=-\frac{C_9}{g_s}\int d^{10}x \str\left(\sqrt{
-\det(E_{\mu\nu}+\lambda F_{\mu\nu})}
\left[\sum_{n=1}^{\infty}f_n(T^2)(
\lambda E^{\mu\nu}
D_{\mu}T D_{\nu}T)^n+V(T) \right]\right) \ ,
\end{equation}
where 
\begin{equation}
\lambda=2\pi\alpha', C_p
=\sqrt{2}T_p, T_p=\frac{2\pi}
{(4\pi^2\alpha')^{(p+1)/2}} 
\end{equation}
and
\begin{equation}
E_{\mu\nu}=G_{\mu\nu}+B_{\mu\nu},
 \ D_{\mu}T=\partial_{\mu}T+i[A_{\mu},T] , \
F_{\mu\nu}=\partial_{\mu}A_{\nu}-\partial_{\nu}A_{\mu}+
i[A_{\mu},A_{\nu}] , \ \mu,\nu=0,\dots,9 \ .
\end{equation}
The gauge field $A_{\mu}$ belongs to the
adjoint representation of the gauge group $U(N)$. Finally,
$V(T)$ is the potential for the tachyon. 
  We do not know much about
 functions  $f_n(T^2)$ with exception that should
be even functions of its argument \cite{SenNA}. There
is also one intriguing conjecture \cite{Garousi,Bergshoff} which
says that
these functions could be equal to the tachyon potential and consequently
should be equal to zero for $T^2=T_{min}^2$, where $T_{min}$ is
a tachyon value at the local minimum. In this paper
we  will suppose that these functions do not need directly equal
to $V(T)$ but we will assume that they are zero for $T=T_{min}$ and
also that they obey $\frac{df_n(T)}{dT}=0$ for $T=T_{min}$.
In other words, we expect these functions in the form
\begin{equation}
f_n(T^2)=\sum_{m=1}b_{nm}(T^2-T_{min}^2)^m \ .
\end{equation}
Then the kinetic term has a form
\begin{equation}
\sum_{n=1}\sum_{m=1}b_{nm}(T^2-T_{min}^2)^m
(\lambda E^{\mu\nu}D_{\mu}TD_{\nu}T)^n \ .
\end{equation}
In (\ref{Action}) the $\str $ means the 
symmetrised trace \cite{Tseytlin}
$ \str(A_1,\dots,A_n)=\frac{1}{n!}(\tr A_1\dots
A_n+ permutations)$. In this trace we consider the field
strength $F$ and covariant derivative $DT$ as a single object
as well as $(T^2-T_{min}^2)$, otherwise we could not obtain
the result that the action is equal to zero for $T=T_{min}$.
The prescription for including the symmetrised trace
was suggested in \cite{Tseytlin}. The evidence for this proposal
was further given in \cite{TaylorM,TaylorM2}. 
 We must also stress one
important thing. It seams to us that  the tachyon potential should
appear as a single object in the action (as for example
a covariant derivative)  in order  to obtain from the
 action the correct value of the tachyon ground state
and also in order to obey the requirement that for the tachyon equal
to its vacuum value the action  should vanish. 
When we used  the potential as a matrix valued function,
than the symmetrised trace would lead to the different arrangements
of the various terms from the tachyon potential and we do not
know how we could get a sensible result.

In order to obtain
the action for lower dimensional D-brane, we use T-duality
in the same manner as in \cite{Garousi,Mayers}. Let us
consider T-duality on a set of directions denoted
with $i,j=p+1,\dots,9$. The fields transform as
\cite{Giveon}
\begin{equation}\label{Trule}
\tilde{E}_{ab}=E_{ab}-
E_{ai}E^{ij}E_{jb}, \
\tilde{E}_{aj}=E_{ak}E^{kj}, 
\ \tilde{E}_{ij}=E^{ij} , \
\end{equation}
where $a,b=0,1,\dots,p$ and $E^{ij}$ denotes
the inverse of $E_{ij}$, i.e., $E_{ik}E^{kj}=
\delta_i^j$. One also has a dilation
transformation
\begin{equation}\label{Tdilaton}
e^{2\tilde{\phi}}=e^{2\phi}\det (
E^{ij}) \ .
\end{equation}
Now T-duality acts to change the dimension
of D-brane world-volume. We have two possibilities:
If a coordinates transverse to Dp-brane, e.g. $y=x^{p+1}$ 
is T-dualised, it becomes D(p+1)-brane where $y$ is
now extra world-volume dimension. If a coordinate
on the world-volume of Dp-brane is T-dualised, e.g.
$y=x^p$, it becomes D(p-1)-brane where $y$ is now
extra transverse dimension. In the first case, we have
a rule for transformation of the world-volume  fields
\begin{equation}
\Phi^{p+1} \Rightarrow A_{p+1} \ ,
\end{equation}
and in the second case
\begin{equation}
A_{y} \Rightarrow \Phi^y \ .
\end{equation}
In the second case, the corresponding field strength
transforms as
\begin{equation}
F_{ay} \Rightarrow D_a\Phi^y \ .
\end{equation}
In the T-duality transformation along the 
world-volume coordinate $x^p$ we presume
that all field are independent on this coordinate
\begin{equation}
\partial_p \Psi =0 \ .
\end{equation}
However, this rule does not imply that $D_{x^p}\Psi$
is equal to zero, rather we obtain
\begin{equation}
D_p\Psi \Rightarrow i[\Phi^p,\Psi ] \ .
\end{equation}
Now we are ready to discuss the action for non-BPS Dp-brane.
We obtain this action from (\ref{Action}) applying T-duality
transformations in $p+1,\dots, 9$ dimensions, following
\cite{Garousi,Mayers}. We will discuss the
term
\begin{equation}
\tilde{D}=\det (\tilde{E}_{\mu\nu}+\lambda F_{\mu\nu}) \ .
\end{equation}
With using (\ref{Trule}) we get
\begin{equation}
D=\det \left(\begin{array}{cc}
E_{ab}-E_{ak}E^{kj}E_{jb}+
\lambda F_{ab} &
E_{ak}E^{kj}+\lambda D_a \Phi^j \nonumber \\
-E^{ik}E_{kb}-\lambda D_b\Phi^i &
E^{ij}+i\lambda [\Phi^i,\Phi^j]
 \nonumber \\
\end{array}\right) \ .
\end{equation}
When we use the  mathematical
formula
\begin{equation}
\det \left (\begin{array}{cc}
A & B \\
C & D \\ \end{array}\right)=
\det \left(\begin{array}{cc}
A-BD^{-1}C & B \\
0 & D \\ \end{array}\right)=
\det( A-BD^{-1}C)\det(D) \ ,
\end{equation}
 we obtain \cite{Mayers}
\begin{equation}
D=\det \left(P\left[E_{ab}+E_{ai}(X^{ij}-\delta^{ij})E_{jb}\right]
\right)\det(Q^{i}_m)\det(E^{mj}) \ ,
\end{equation}
where we have defined
\begin{equation}
Q^{ij}=E^{ij}+i\lambda [\Phi^i,\Phi^j] \ ,
P[E_{ab}]=E_{ab}+2\lambda E_{ai}D_b\Phi^i+
\lambda^2E_{ij}D_a\Phi^iD_b\Phi^j \ ,
\end{equation}
and
\begin{equation}
X^{kl}=E^{ki}(Q)^{-1}_{ij}E^{jl} \ .
\end{equation}
We have also used
\begin{equation}
\det (Q^{ij})=\det(Q^{im}E_{mk}
E^{kj})=\det (Q^i_m)\det(E^{mj}) \ .
\end{equation}
The analysis of $F$ function in (\ref{Action}) is
straightforward and we get
\begin{eqnarray}\label{Ftdual}
F=V(T)-\sum_{n=1}^{\infty}f_n(T)\lambda^n
\left((E^{ab}-E^{ai}E_{ij}E^{jb})D_aTD_bT
+\right.\nonumber \\
\left. +2iE^{ak}E_{kj}[\Phi^j,T]D_aT-
E_{ij}[\Phi^i,T][\Phi^j,T]\right)^n  \ .\nonumber 
\end{eqnarray}
With using (\ref{Tdilaton}) we obtain the action
for non-BPS Dp-brane 
\begin{eqnarray}\label{action2}
S=-\frac{C_p}{g_s}\int d^{p+1}\sigma\str\left(\sqrt{-
\det(P[E_{ab}+E_{ai}(X^{ij}-\delta^{ij})
E_{jb}]}\sqrt{\det Q^i_j}F(T,DT,\dots)\right) \ . \nonumber \\
\end{eqnarray}
\section{Applications}\label{third}

In this section we will discuss the various applications 
of the previous action.
We will  work with the non-abelian action for $N$ 
non-BPS D0-branes in Type IIB theory. Thanks to gauge invariance, we can
pose $A_0=0$. Than the covariant derivative is $D_t\Phi=
\dot{\Phi}$. We will work in the flat space-time
background 
\begin{equation}
E_{ab}=\eta_{ab}, \  a,b=0, \ 
E^{ij}=\delta^{ij}, \  i,j=1,\dots,9 \ .
\end{equation}
Then we have
\begin{equation}
Q^{ij}=\delta^{ij}+i\lambda [\Phi^i,\Phi^j] \ ,
\end{equation}
\begin{equation}
P[E_{ab}]=-1+\lambda^2(\dot{\Phi}^i)^2 \ ,
\end{equation}
\begin{equation}
P[E_{ai}X^{ij}E_{jb}]=\lambda^2\dot{\Phi}^k
\delta_{ki}X^{ij}\delta_{jl}\dot{\Phi}^l \ , 
\end{equation}
and finally
\begin{equation}\label{PE}
P[E_{ab}+E_{ai}(X^{ij}-\delta^{ij})E_{jb}]=
-1+\lambda^2(\dot{\Phi}^i)^2+\lambda^2
\dot{\Phi}^k\delta_{ki}(X^{ij}-\delta^{ij})\delta_{jl}
\dot{\Phi}^l \ .
\end{equation}
We will work in the leading order in $\lambda$ in
which the previous expression reduces into
\begin{equation}
-1+\lambda^2(\dot{\Phi}^i)^2 
\end{equation}
 and $F(T,\dots)$ in 
the leading order approximation
 has a form 
\begin{equation}\label{TD0}
F=V(T)+f(T)\lambda \dot{T}\dot{T}
+\lambda f(T)\delta_{ij}[\Phi^i,T][\Phi^j,T] \ .
\end{equation}
Using these results we 
 obtain the action for $N$ non-BPS D0-branes
in  the leading order approximation
\begin{eqnarray}\label{actD0}
S=\frac{C_0}{g_s}\int dt
\str \left(-V(T)+\frac{\lambda}{2}\dot{\Phi}^i
\dot{\Phi}^j\delta_{ij}-\frac{\lambda^2}{4}
\delta_{kl}\delta_{mi}[\Phi^i,\Phi^k]
[\Phi^l,\Phi^m]V(T) \right. + \nonumber \\
\left. +\lambda f(T)\dot{T}\dot{T}+
\lambda f(T)\delta_{ij}[\Phi^i,T][\Phi^j,T] \right) \ .
\nonumber \\
\end{eqnarray}
In what follows we will consider static
configurations only, when all fields are time independent. 
Then the action is
\begin{eqnarray}\label{Npot}
S=-\frac{C_0}{g_s}\int dt  \mathcal{V}(T,\Phi) \ , \nonumber \\
\mathcal{V}(T,\Phi)=\str \left(
V(T)+\frac{\lambda^2}{4}[\Phi^i,\Phi^j]
[\Phi^j,\Phi^i]V(T)-\lambda f(T)[\Phi^i,T]
[\Phi^i,T] \right) \ . \nonumber \\
\end{eqnarray}
We will also consider the coupling of the non-BPS
D-brane to the external RR field. This term was
calculated in \cite{Billo} for single non-BPS
D-brane and we have generalised this term
for $N$ non-BPS D-branes in \cite{KlusonP1,KlusonP2}.
Applying T-duality rules on these terms give
 complicated expression which was discussed 
recently in \cite{Janssen}. However in this
paper we will discuss only the leading order term 
which for non-BPS D0-branes has a form
\begin{equation}\label{WZcal}
\frac{\mu_{-1}}{2T_{min}}\int dt \str P\left[ 
e^{i\lambda\I_{\Phi}\I_{\Phi}}\left(\dot{T}+i[\I_{\Phi},T]\right)
\sum_n C^{(n)}\right] \ ,
\end{equation}
where $P$ is a pull-back to the world-volume of
D0-brane and $\I_{\Phi}$ is the interior product
\cite{Nakahara}. Acting on forms, the interior product
is an anticommuting operator of form degree $-1$, e. g.,
\begin{eqnarray}
C^{(2)}=\frac{1}{2}C_{\mu\nu}^{(2)}
dx^{\mu}\wedge dx^{\nu} \ , \nonumber \\
\I_{v}C^{(2)}=v^{\mu}C_{\mu\nu}^{(2)}dx^{\nu} \ ,
\nonumber \\
\I_{w}\I_{v}=w^{\nu}v^{\mu}C_{\mu\nu}^{(2)}=
-\I_{v}\I_{w}C^{(2)} \ .
\end{eqnarray}
We will see that this Wess-Zumino term (WZ) correctly
reproduces the charges of higher dimensional D-branes
arising from tachyon condensation on the system
of $N$ non-BPS D0-branes. 

We start with  simple examples of tachyon condensation
which are the solutions of the whole action as well. The first one is
\cite{Harvey2}
\begin{equation}
T=T_{min}(1-P_k)=
\mathrm{diag} (0,\dots, 0^k, T_{min},\dots,T_{min}^{N-k}), \
\Phi^i=0 \ , 
\end{equation}
where $P_k$ is a projector on the fist $k$ states which has 
 the form $P_k=\mathrm{diag}(1,\dots,1^k)$.
It is easy to see that this is a solution of the equation
of motion  since all commutators vanish (we do not need
presume the condition $f_n(T_{min})=0$) and also it is easy
to see that $\frac{dV}{dT}=0$. The energy of this configuration
is equal to 
\begin{equation}
E=\frac{C_0}{g_s}\str V(T)=\frac{C_0}{g_s}\tr V(T)=
\frac{C_0}{g_s}k \ ,
\end{equation}
where we have used $V(0)=1$ \cite{SenFT1}. This 
is the  rest energy of $k$ non-BPS D0-branes. However,
there is also one nontrivial solution 
corresponding to tensionless D0-branes
\cite{Harvey2}
\begin{equation}\label{tensionless}
T=T_{min}(1-2P_k)=
\mathrm{diag}(-T_{min},\dots,-T_{min}^k,T_{min},
\dots, T_{min}^{N-k}), \ \Phi^i=0 \ .
\end{equation}
As in the previous solution commutators are equal to zero
and the variation of the potential gives 
\begin{equation}
\frac{\delta V(T)}{\delta T}=
\sum_{n=1}^{\infty} n a_n 2T(T^2-T_{min}^2)^{n-1} \ ,
\end{equation}
since we can presume that the potential has a form
$V(T)=\sum_{n=1}^{\infty}a_n (T^2-T_{min}^2)^n$. 
Then we can immediately see that (\ref{tensionless}) is a solution 
which
is a extreme of the potential since
\begin{equation}
T_{min}^2(1-2P_k)^2=T_{min}^2 \ .
\end{equation}
The energy of this configuration is equal to
\begin{equation}
E=\frac{C_0}{g_s}\tr V(T)=0 \ ,
\end{equation}
since $T^2=T_{min}^21_{N\times N}$.
What is a physical meaning of this object? We think
that this object is equivalent to the tensionless D0-branes
discovered recently in \cite{Gopakumar, Harvey} in the
framework of noncommutative theory. It was argued in
\cite{Harvey2} that these objects are gauge equivalent
to the vacuum.  The same problem was discussed in
\cite{WittenFT}. We see that we can obtain tensionless
D0-brane in our approach. It would be very interesting to
study fluctuation around this solution. We hope to return
to this puzzle in the future.

\section{D1-brane}\label{fourth}
In this section we will consider more general
solution when $V(T)$ and $f(T)$ does not
commute with $\Phi$. 
Then the equation of motion for $\Phi^i$ 
 has a form
\begin{eqnarray}\label{equationX2}
 \lambda [T,[\Phi^i,T]f(T)]+
\lambda[T,f(T)[\Phi^i,T]]
+\nonumber \\
+\frac{\lambda^2}{2}[\Phi^k,[\Phi^i,\Phi^k]V(T)]+
\frac{\lambda^2}{2}[\Phi^k,V(T)[\Phi^i,\Phi^k]]=0 \ ,\nonumber \\
\end{eqnarray}
and the equation of motion for  tachyon
\begin{eqnarray}\label{equationTf2}
\frac{dV(T)}{dT}\left(1-
\frac{\lambda^2}{4}[\Phi^i,\Phi^j]
[\Phi^i,\Phi^j]\right)-\lambda
\frac{df(T)}{dT}
[\Phi^i,T][\Phi^i,T]
-\nonumber \\
-\lambda\left(
[[\Phi^i,T]f(T),\Phi^i]+
[f(T)[\Phi^i,T],\Phi^i]\right)=0 \ . \nonumber \\
\end{eqnarray}
We take an ansatz
\begin{equation}\label{clas}
T=F(\hat{x}_1)=\sum b_n \hat{x}_1^n,  \ \Phi^2=
k^{-1}\hat{x}_2, \  
  [\hat{x}_1,\hat{x}_2]=ik, \
\Phi^i=0, \ i=1,3,\dots,9 \ ,
\end{equation}
where $F(x)$ approaches $T_{min}$ for
$x\rightarrow -\infty$ and $-T_{min}$ for
$x\rightarrow \infty$.
Then
\begin{equation}
[\hat{x}^2_1,\hat{x}_2]=2ik\hat{x}_1 , \ 
[\hat{x}_1^3,\hat{x}_2]=3ik\hat{x}_1^2,\
\dots, \ [\hat{x}_1^{n},\hat{x}_2]=
nik \hat{x}_1^{n-1} \ .
\end{equation}
Using this result we obtain
\begin{equation}\label{tachderivative}
[T,\Phi^2]=k^{-1}[\sum_{n}b_n\hat{x}_1^n,
\hat{x}_2]
=i\sum_n b_n \hat{x}_1^{n-1}n
=i\frac{dT}{d\hat{x}_1} \ , 
\end{equation}
and consequently
\begin{equation}\label{tachderivative2}
[T^2,\Phi^2]=2iT\frac{dT}{d\hat{x}_1} \ ,
[T^4,\Phi^2]=4iT^3\frac{dT}{d\hat{x}_1},
\ \dots, \ [T^{2n},\Phi^2]=i 2n T^{2n-1}
\frac{dT}{d\hat{x}_1} \ .
\end{equation}
With these results in hand we obtain
\begin{equation}
[[\Phi^2,T]f(T),\Phi^2]=\frac{d}{d\hat{x}_1}
(T' f(T))
\ ,
[f(T)[\Phi^2,T],\Phi^2]=\frac{d}{d\hat{x}_1}
(T'f(T)) \ ,
\end{equation}
where $T'=\frac{dT}{d\hat{x}_1}$. 
Then the equation of motion for tachyon has
a form
\begin{equation}\label{xx}
\frac{dV}{dT}
-\lambda\frac{df}{dT}T'^2
-2\lambda fT''=0 \ .
\end{equation}
After multiplication with $T'$ we get the result
\begin{equation}\label{garousi}
V'(T(\hat{x}_1))
=(\lambda  fT'^2)'
\rightarrow V(T)=\lambda f(T)T'^2 \ ,
\end{equation}
where the integration constant has been set to
zero. In the previous expression we have used
\begin{equation}\label{pp}
(V(T(\hat{x}_1)))'=\sum a_n (T^{2n})'=
\sum a_n 2nT^{2n-1}T'=\frac{dV}{dT}T' \ .
\end{equation}
The equation of motion for $\Phi^2$ is trivially satisfied
since
\begin{equation}
[T,[\Phi^2,T]f(T)]=-i[T,T'V(T)]=0 \ .
\end{equation}
An energy of this solution is 
\begin{equation}\label{erg}
E=\frac{C_0}{g_s}\str\mathcal{V}(T)=
2\frac{C_0}{g_s}\tr V(T(\hat{x}_1)) \ , 
\end{equation}
where we have used (\ref{garousi}).
Since we do not known the exact form of the $f(T)$ function
we cannot determine the tachyon field so that
 we will work with (\ref{erg})
without exact form of the tachyon field $T=
F(\hat{x}_1)$. Note that in this solution we do not
need to presume that $T_{min}$ is
extreme of $f(T)$ with $f(T_{min})=0$. It seams that
this solution is more general one than the solution given
in the section (\ref{third}).

We must stress one important thing. We work
 in this section in the limit $N\rightarrow \infty$,
when we can replace the matrices with the abstract
operators action on Hilbert space, in the same
way as in Matrix theory \cite{BanksM,Banks,Taylor,Taylor2}.
Then $\hat{x}_1,\hat{x}_2$ are as the same operators
as operators of coordinate and impuls for one particle
system in standard quantum
mechanics and we can easily calculate the trace
in (\ref{erg}) 
\begin{eqnarray}
E=\frac{C_0}{g_s}\tr 2V(T(\hat{x}_1))
=2\frac{C_0}{g_s}\int dx_2\bra{x_2}
V(T(\hat{x}_1))\ket{x_2}=\nonumber \\
=\frac{2C_0}{g_s}\int dx_2dx'_1dx''_1
\left<x_2|x'_1\right>V(T(x_1))
\left<x'_1|x''_1\right>\left<x''_1|x_2\right>=
\nonumber \\
=2\frac{C_0}{g_s}\int dx_2dx_1
|\left<x_2|x_1\right>|^2 V(T(x_1))=
\frac{2C_0}{g_s2\pi k}
\int dx_1dx_2 V(T(x_1)) \ , \nonumber \\
\end{eqnarray}
where we have used the standard normalisation
\begin{equation}
\left<x_1|x_2\right>=\frac{1}{\sqrt{2\pi k}}e^{ix_1x_2/k} \ ,
\end{equation}
where $\ket{x_1},\ket{x_2}$ are eigenvectors of 
$\hat{x}_1$ and $\hat{x}_2$  with the normalisation
condition $\left<x_1|x'_1\right>=\delta (x_1-x'_1),
\left<x_2|x'_2\right>=\delta (x_2-x'_2)$.
In  \cite{SenFT3,Kluson} 
the energy of tachyon lump on unstable non-BPS D-brane
was calculated. It was shown that the tension of
resulting lump (in linearised approximation)
 is given with the integral
$C_0 \int dx V(T(x))\sim T_{-1}=2\pi$. We cannot write
equality since we do not know the precise form
of $T=F(x)$ and we do not know the exact form o tachyon
potential. However we can expect that this integral
gives the result proportional to the tension of D(-1)-brane and
then we get
\begin{equation}
E\sim\frac{2\pi}{g_s4\pi^2\alpha' \tilde{k}g_s}\int dx_2=
\frac{T_1}{g_s \tilde{k}} \int dx_2 \ , \tilde{k}=k\lambda^{-1} \ ,
\end{equation}
which corresponds to the energy of  D1-brane. 
The factor $\tilde{k}$ can be absorbed with coordinate
redefinition. We claim that  the energy of this configuration
corresponds to the energy of D1-brane. This conclusion
is also supported with the analysis of the WZ term (\ref{WZcal})
\\
\begin{eqnarray}\label{charges}
I_{WZ}=\frac{\mu_{-1}}{
2T_{min}}\int dt \tr i[\Phi^2,T]C_{2t}^{(2)}
=\frac{\mu_{-1}}{
2T_{min}}\frac{1}{2\pi k}
\int dt dx_1 dx_2  \frac{dT(x_1)}{dx_1}C_{2t}^{(2)}=\nonumber \\
=\frac{\mu_{-1}}{
2T_{min}}\frac{1}{4\pi^2 \alpha'\tilde{k}}
\int dt dx_1 \left(T(\infty)-T(-\infty)\right)C_{2t}^{(2)}
=\mu_1\int dt dx_2 C_{t2}^{(2)} \ , \nonumber \\
\end{eqnarray} 
where we have used $ T(\infty)=-T_{min}, 
T(\infty)=T_{min}$, and have dropped the factor $\tilde{k}$. 
This is precisely the coupling between
D1-brane and RR two form. It is remarkable that
this exact result  does not depend on the precise form
of tachyon field. We hope that this result
gives strong evidence that (\ref{clas}) really
leads to the emergence of D1-brane. 
However,  we must stress again  that it seams 
to be hopeless to calculate exactly the
energy of resulting configuration without
knowledge of exact BI action for non-BPS
D0-brane. On the other hand, recent results
given in \cite{SenTP} suggest that higher
derivative terms could be gauge artefacts only
and then it seams to be possible to obtain
exact solution.

\subsection{Other BPS D-branes from
non-BPS D0-branes}

In this subsection we will see that we
can obtain all BPS D-branes through
tachyon condensation in the same way
as a D1-brane. Let us consider an ansatz
\begin{eqnarray}\label{clas2}
T=F(\hat{x}_1),  \Phi^1=
k^{-1}\hat{x}_2, \  
 [\hat{x}_1,\hat{x}_2]=ik \ , \nonumber \\
\Phi^{2i}=k_i^{-1}\hat{p}_{i},\
\Phi^{2i+1}=k_i^{-1}\hat{q}_{i}, \
[\hat{p}_{i},\hat{q}_{i}]=ik_i, \  i=1,\dots,p \ ,
\nonumber \\
\Phi^i=0, \  i=2p+2,\dots, 9 \ . \nonumber \\
\end{eqnarray}
It is easy to see that this ansatz solves
 (\ref{equationX2})  and
 from (\ref{equationTf2}) we obtain
\begin{equation}
V(T(\hat{x}_1))(1+\frac{\lambda^2}{2}
\sum_{i=1}^p\frac{1}{k_i})=\lambda 
f(T(\hat{x}_1)) T'^2 \ .
\end{equation}
We choose the Hilbert space basis 
\begin{equation}
\ket{\psi}=\ket{x_1}\otimes
\ket{p_1}\otimes \dots \otimes\ket{p_p}
\ ,
\left<\psi|\psi'\right>=\delta (x_1-x_1')
\delta (p_1-p_1')\dots \delta (p_p-p'_p) \ .
\end{equation}
Then the energy of given configuration is equal to
\begin{equation}
E=\frac{C_0}{g_s}2\tr V(T(x_1))(1+\frac{\lambda^2}
{2}\sum_{i=1}^p\frac{1}{k_i^2})\sim
\frac{T_{2p+1}}{\tilde{k}\prod_{i=1}^p
\tilde{k}_i}(1+\frac{1}{2}
\sum_{i=1}^p\frac{1}{\tilde{k}_i^2})\int dx_2dp_1dq_1
\dots dp_pdq_p \ .
\end{equation}
This is proportional to the 
 energy of D(2p+1)-brane since this energy scales
as $V_{2p+1}$ which is a volume on which this D-brane
is wrapped. We have also defined $\tilde{k}_i=\lambda^{-1}
k_i$. These factors can be dropped out from the action after
 redefinition $dp_i dq_i/\tilde{k}_i\rightarrow dx_{2i}dx_{2i+1}$.
Since we used the action in linear approximation, the 
commutators should obey
\begin{equation}
\lambda^2 [\Phi^i,\Phi^k]^2\ll 1 \Rightarrow
\frac{\lambda}{k}\ll 1 \ . 
\end{equation}
In  limit $k\rightarrow \infty$ we can neglect the
second term in the bracket and after the second
redefinition $x_2\rightarrow x_1 $ we obtain
the result 
\begin{equation}
E\sim T_{2p+1}\int dx_1 dx_2\dots dx_{2p+1} \ ,
\end{equation}
which suggests that the resulting configuration
is  D(2p+1)-brane. 
This claim  is also supported with the analysis of the Wess-Zumino
term which has a form 
\begin{eqnarray}\label{chargegeneral}
I_{WZ}=\frac{\mu_{-1}}
{2T_{min}}\int dt \str \left(
e^{i\lambda \I_{\Phi}\I_{\Phi}}
i[\I_{\Phi},T] \sum_n C^{(n)}\right)=
\frac{\mu_{-1}}{2T_{min}}\int dt \str i[\Phi^1,T]
C_{1t}^{(2)}-\nonumber \\
-\frac{\mu_{-1}}{4T_{min}}\sum_{i=1}^p
\int dt \str  \lambda [\Phi^{2i},\Phi^{2i+1}]
[\Phi^1,T]C_{1,2i+1,2i,t}^{(4)}- \nonumber \\
-\frac{i\mu_{-1}}{8T_{min}}
\sum_{i=1,j\neq i}^p
\int dt\str\lambda^2 [\Phi^{2i},
\Phi^{2i+1}][\Phi^{2j},\Phi^{2j+1}]
[\Phi^1,T]C_{1,2j+1,2j,2i+1,2i,t}^{(6)}+  \nonumber \\
\dots+
\frac{i(i)^p\lambda^p\mu_{-1}}{2T_{min}}
\int dt \str [\Phi^{2},\Phi^3]\dots [
\Phi^{2p},\Phi^{2p+1}][\Phi^1,T]
C_{1,2p+1,2p,\dots,3,2,t}^{(2p+2)} \ .\nonumber \\
\end{eqnarray}
The first term in (\ref{chargegeneral}) corresponds to
the  coupling of D1-brane 
to two form  $C^{(2)}$, as we have seen in
(\ref{charges}). We will see that this configuration is
 charged with respect to  $C^{(2)},C^{(2)},\dots,
C^{(2p)}$ 
\footnote{Since the various commutators are pure
numbers we can replace the symmetrised trace with
the ordinary one.}. The same thing also arises in the
 construction of higher dimensional objects in
Matrix theory \cite{BanksM,Banks,
Taylor,BanksR1,BanksR2,Taylor2}. The second term in 
(\ref{chargegeneral}) gives
\begin{equation}
\sum_{i=1}^p \frac{\mu_{1}}{2\pi \lambda \tilde{k}_i}
\int dt dx_1 dp_{i}dq_{i}C_{t1,2i,2i+1}^{(4)}=
\sum_{i=1}^p\mu_3
\int dt dx_1 dx_{2i}dx_{2i+1}C_{t1,2i,2i+1}^{(4)}=
\sum_{i=1}^p\mu_3\int_{M_i} C^{(4)} \ ,
\end{equation}
where $M_i$ is  a submanifold parameterised with $t,x_1,x_{2i},
x_{2i+1}$. It is clear that the previous term corresponds to
$p$ D3-branes wrapped submanifolds $M_i$. Again, their
interpretation is the same as in Matrix theory. 

 In the same
way we can proceed with other terms. For example, let us
consider the third term in (\ref{chargegeneral}) with $i=1,j=2$.
Then we obtain
\begin{eqnarray}
-\frac{i\mu_{-1}\lambda^2}{2T_{min}}
\int dt \str [\Phi^2,\Phi^3][\Phi^4,\Phi^5][\Phi^1,T]
C^{(6)}_{15432t}=-\frac{\mu_{5}}{\tilde{k}_1^2
\tilde{k}_2^2\tilde{k}}\int dtdx_2
dp_1dq_1dp_2dq_2 C_{15432t}=\nonumber \\
=\mu_{5}\int dt dx_1 dx_2 \dots dx_5 C_{t1\dots 5}=
\mu_{6}\int_{M_{12}} C^{(6)} \ , \nonumber \\
\end{eqnarray}
 where $M_{12}$ is a six dimensional submanifold parameterised
with coordinates $t,x_1,\dots, x_5$. 
Finally, the last term in (\ref{chargegeneral}) gives
\begin{equation}
\mu_{2p+1}\int dt dx_1 dx_2 \dots dx_{2p} C_{t12\dots 2p}
^{(2p+2)}=\mu_{2p+1}\int C^{(2p+2)} \ . 
\end{equation}
We see that this configuration really corresponds to
the BPS D(2p+1)-brane.  The
fact that $\Phi^i$ in (\ref{clas2}) do not commute suggests
that the world-volume of resulting configuration
is noncommutative. It would be nice to study
the fluctuation around this static solution. We 
hope to return to this interesting question in the future.

In the next section we would like to show that the
action for $N$ non-BPS D0-branes naturally leads to
the non-commutative action for any higher dimensional
non-BPS D-brane, following \cite{Seiberg}.

\section{Non-BPS D-branes from 
non-BPS D0-branes}\label{fifth}
We have seen in the (\ref{third}) section that the action
for $N$ non-BPS D0-branes has a trivial solution
\begin{equation}
T=T_{min} 1_{N\times N} , \Phi^i=0
\ , i=1,\dots,9 \ ,
\end{equation}
corresponding to the unstable vacuum with $N$ non-BPS
D0-branes. 
There is a question whether we can construct  
other  non-BPS D-branes. Let us answer this
question, following \cite{Seiberg} and also earlier works
\cite{Banks,Li,Aoki,Ishibashi,Ishibashi1}.

We start with the action
\begin{eqnarray}\label{action3}
S=-\frac{C_0}{g_s}
\int dt\str\left(\sqrt{-
\det(P[E_{tt}+E_{tI}(X^{IJ}-\delta^{IJ})
E_{Jt}]}\right)\times\nonumber \\
\times\sqrt{\det (Q^I_J})\times F(T,\dot{T},\Phi^I,\dots) \ , \nonumber \\
\end{eqnarray}
with 
\begin{eqnarray}\label{Ftdual2}
F=V(T)-\sum_{n=1}^{\infty}f_n(T)\lambda^n
\left((E^{tt}-E^{tI}E_{IJ}E^{Jt})\dot{T}\dot{T}
+\right.\nonumber \\
\left.+iE^{tK}E_{KJ}[\Phi^J,T]\dot{T}-
E_{IJ}[\Phi^I,T][\Phi^J,T]\right)^n \ .
\nonumber \\
\end{eqnarray}
We
introduce the constant  background metric $E_{IJ}=g_{IJ}$,
with vanishing $B_{IJ}$ and with $E^{tI}=0$. 
Then $Q^{IJ}=g^{IJ}+i\lambda[\Phi^I,\Phi^J]$.)
We also use the notation $I,J,K,\dots=1,\dots, 9 ; 
i,j,k,\dots=1,\dots, 2p ; a,b,c,\dots=2p+1,\dots,9$.
We also assume that this background metric is block-diagonal
with the blocks $g_{ij},g_{ab}$ with $g_{ia}=0$.
Let us propose an ansatz
\begin{equation}\label{ansatznBPS}
T=0 \ ,
\Phi^i_{clas}=\lambda^{-1} x^i, \  i=1,\dots, 2p, \ 
[x^i,x^j]=i\Theta^{ij}
\end{equation}
and  other fields $\Phi^{a}, \  a=2p+1,
\dots, 9$ in the form $\Phi^a=x^a 1_{N\times N}$,
which describe the transverse positions of the resulting
D-brane. It is easy to see that this ansatz (\ref{ansatznBPS})
 is a solution of the equation of motion of the whole action 
 since the commutators between tachyon and scalar field
vanish and also from the fact that commutators of two $\Phi's$ are
pure numbers and consequently $[\Phi^i,[\Phi^j,\Phi^i]]=0$.

Next we will analyse the fluctuation around this background.
We will closely follow \cite{Seiberg} and write
\begin{eqnarray}\label{ansatz}
C_i=\lambda B_{ij}\Phi^j=\lambda B_{ij}\Phi_{clas}^j+\lambda \Phi^j_{fluct}=
 B_{ij}x^j+\hat{A}_i, \  i=1,\dots ,2p \ , \nonumber \\
\Phi^a=\Phi^a,  \  a=2p+1,\dots, 9, \ 
 T_{fluct}=T \ , \nonumber \\
\end{eqnarray}
and calculate
\begin{eqnarray}\label{CC}
[C_i,C_j]=-iB_{ij}+
B_{ik}[x^k,\hat{A}_j]-B_{jl}[x^l,\hat{A}_i]+
[\hat{A}_i,\hat{A}_j]  \ , \nonumber \\
\end{eqnarray}
\begin{equation}
[C_i,\Phi^a]=B_{ik}[x^k,\Phi^a]+
[\hat{A}_i,\Phi^a] \ ,
\end{equation}
where we have used
\begin{equation}
B_{ik}B_{jl}[x^k,x^l]=
-B_{ik}i(B^{-1})^{kl}B_{lj}=-iB_{ij} \ .
\end{equation}
Using
\begin{equation}
\Phi^i=\lambda^{-1}\Theta^{ik}C_k \ ,
\end{equation}
we obtain
\begin{equation}
i\lambda [\Phi^i,\Phi^j]
=\lambda^{-1}\Theta^{ik}(
\hat{F}_{kl}-B_{kl})\Theta^{lj} \ ,
\end{equation}
where
\begin{equation}
\hat{F}_{kl}=-iB_{ik}[x^k,\hat{A}_l]+
iB_{jl}[x^l,\hat{A}_i]-i[\hat{A}_i,\hat{A}_j] \ .
\end{equation}
The best thing how we can study the fluctuation
around the classical solution is to start with the original
form of the determinant
\begin{equation}
D=\det \left(\begin{array}{cc}
g_{tt}&\lambda D_{t}\Phi^J \nonumber \\
-\lambda D_t\Phi^I&
g^{IJ}+i\lambda [\Phi^I,\Phi^J]
 \nonumber \\
\end{array}\right)=
\det \left(\begin{array}{ccc}
D_{tt} & D_{tj} & D_{tb} \\
D_{it} & D_{ij} & D_{ib} \\
D_{at} & D_{aj} & D_{ab} \\ \end{array}\right) \ ,
\end{equation}
with the action in the form
\begin{equation}\label{actionD}
S=-\frac{C_0}{g_s}\int dt
\str \sqrt{\det g_{IJ}}\sqrt{-\det D}\times F(T,\Phi,\dots)
\ , 
\end{equation}
where the factor $\sqrt{\det g_{IJ}}$ arises from T-duality
transformation of the dilation (\ref{Tdilaton}).

We have also  written $D_t$ instead $\partial_t$
in order to obtain more symmetric expression. (Remember,
we are working in gauge $A_0=0$.) 
Then we obtain
\begin{equation}
D_{tb}=\lambda D_t\Phi^b \ , 
D_{it}=-\Theta^{ik} D_t\hat{A}_k \ ,
D_{tj}= -D_t\hat{A}_k\Theta^{kj} \ ,
D_{at}=-\lambda D_t\Phi^a \ ,
\end{equation}
\begin{equation}
D_{ij}=g^{ij}+i\lambda[\Phi^i,\Phi^j]
=g^{ij}+\lambda^{-1}\Theta^{ik}(
 \hat{F}_{kl}-B_{kl})\Theta^{lj}\ , 
\end{equation}
\begin{equation}
D_{ib}=-\Theta^{ik}D_k\Phi^b \ ,
D_{aj}=-D_k \Phi^a\Theta^{kj} \ ,
\end{equation}
where
\begin{equation}
iD_kM=[C_k,M]=B_{kl}[x^l,M]+
[\hat{A}_l,M] \ .
\end{equation}
Finally we have
\begin{equation}
D_{ab}=g^{ab}+i\lambda [\Phi^a,\Phi^b]
=Q^{ab} \ .
\end{equation}
Then
\begin{equation}
D=\det\left(
\begin{array}{ccc}
D_{tt}-D_{tb}(Q^{-1})_{bc} D_{ct} &
D_{tj}-D_{tb}(Q^{-1})_{bc}D_{cj} & D_{tb} \\
D_{it}-D_{ib}(Q^{-1})_{bc}D_{ct} &
D_{ij}-D_{ib}(Q^{-1})_{bc}D_{cj} & D_{ib} \\
0 & 0 & Q^{ab} \\ \end{array}\right)
\end{equation}
The first block-diagonal term suggests emergence
of D(2p)-brane. We will show this more precisely
\begin{equation}
D_{tt}-D_{tb}(Q^{-1})_{bc}D_{ct}=
g_{tt}+\lambda^2D_t\Phi^a (Q^{-1})_{ab}D_t\Phi^b
=P[g_{tt}+g_{ta}(X^{ab}-\delta^{ab})
g_{bt}] \ ,
\end{equation}
where  meaning of various terms it the same as
in  the section (\ref{third}). In the similar way we obtain
\begin{eqnarray}
D_{tj}-D_{tb}(Q^{-1})_{bc}D_{cj}=
(-D_t\hat{A}_k+\lambda D_t\Phi^b
(Q^{-1})_{bc}D_k\Phi^c)\Theta^{kj} 
=\nonumber \\
=(-\lambda \hat{F}_{tk}+
P[g_{tk}+g_{ta}(X^{ab}-\delta^{ab})g_{bk}]
)\lambda^{-1}\Theta^{kj} \ , \nonumber \\
D_{it}=-\lambda^{-1}\Theta^{ik}
 (-\lambda \hat{F}_{kt}
+P[g_{kt}+g_{ka}(X^{ab}-\delta^{ab})g_{bt}]) \ ,
\nonumber \\
\end{eqnarray}
\begin{eqnarray}
D_{ij}-D_{ib}(Q^{-1})_{bc}
D_{cj}=g^{ij}-\lambda^{-1}(\Theta B \Theta)^{ij}
+\lambda^{-1}(\Theta \hat{F}\Theta)^{ij}-\Theta^{ik}
D_k\Phi^b(Q^{-1})_{bc}D_b\Phi^c\Theta^{kj}
=\nonumber \\
=-\Theta^{ik}\lambda^{-1}(B_{kl}-
\hat{F}_{kl}+
P[G_{kl}+G_{ka}(X^{ab}-\delta^{ab})G_{bl}])
\Theta^{lj} \ ; 
G_{ij}=-\lambda^2\Theta_{ik}g^{kl}\Theta_{lj} \ .
\nonumber\\
\end{eqnarray}
We combine  $D_{tt}$  with $D_{ij},D_{it},D_{ij}$
into one single matrix $\mathcal{D}_{ij},
i,j=0,1,\dots,2p$. Then $D$ is equal to
\begin{equation}
D=(\det \lambda \Theta)^2\det \mathcal{D}'\det Q^{ij} \ ,
\mathcal{D}=\Theta \mathcal{D}'\Theta \ ,
\end{equation} 
where we have used
\begin{equation}
\mathcal{D}=\left(\begin{array}{cc}
A & BX \\
YC & -YPX \\ \end{array}\right)=
\left(\begin{array}{cc}
1 & 0 \\
0 & -Y \\ \end{array}\right)
\left(\begin{array}{cc}
A & B \\
-C & P \\ \end{array}\right)
\left(\begin{array}{cc}
1 & 0 \\
0 & X \\ \end{array}\right) \ .
\end{equation}
In previous expression $X=Y=\lambda \Theta$.
We can also write
\begin{equation}
\det{g}\det{Q^{ab}}=\det(Q^a_b) \ ,
\end{equation}
where we have used the fact that  the original action (\ref{actionD}) 
contains the factor $\sqrt{\det(g_{IJ})}=\sqrt{\det(g_{ij})}\sqrt{\det
(g_{ab})}$.
We can analyse  $F$ function (\ref{Ftdual2}) in  the similar way.
More precisely
\begin{equation}
-\lambda^{-1}\Theta^{ik}
\Theta^{jl}[C_k,T][C_l,T]=
\lambda^{-1}\Theta^{ik}
\Theta^{jl}D_kTD_lT=
-\lambda^{-1}(\Theta DT DT\Theta)^{ij} \ ,
\end{equation}
and consequently
\begin{eqnarray}
-g_{ij}[\Phi^i,T][\Phi^j,T]=
-g_{ij}\lambda^{2}[\Theta^{ik}C_k,T]
[\Theta^{jl}C_l,T]=\nonumber \\
=-G^{kl}[C_k,T][C_l,T]=G^{ij}D_iTD_jT \ . \nonumber \\
\end{eqnarray}
As a result we obtain
\begin{equation}
F=V(T)-\sum_{n=1}f_n(T)\lambda^n\left(
G^{ij}D_iTD_jT-g_{ab}[\Phi^a,T][\Phi^b,T]\right)^n \ ,
\end{equation}
where we have included $g_{tt}$ into the definition of $G_{ij}$.

With these results in hand we get the final result
\begin{eqnarray}\label{actionfinal}
S=-\frac{C_0}{g_s}\str\int dt
\sqrt{\det(g_{ij})}\det(\lambda^{-1} \Theta)
\times \nonumber \\
\times\sqrt{-\det(\lambda
(\hat{F}_{kl}-B_{kl})+
P[G_{kl}+G_{ka}(X^{ab}-\delta^{ab})G_{bl}]}
\sqrt{\det(Q^a_b)}\times
F(T,DT,\dots) \ , \nonumber \\
\end{eqnarray}
where we have used
\begin{equation}
\det(A+B)=\det(A+B)^T=
\det(A-B) \ , A^T=A, B^T=-B \ .
\end{equation}
In the previous equation the trace goes over $N\times N$ matrices.
Following \cite{Seiberg}, we can take the  limit
$N\rightarrow \infty$. Then there is a standard relation between
the trace over Hilbert space and integration in noncommutative
theory, see \cite{StromingerA,Seiberg}
\begin{equation}
\int d^{2p}x=(2\pi)^n\sqrt{\det{\Theta}}\tr \ .
\end{equation}
We must also remember that  the multiplication in the
resulting action  is noncommutative one with 
the ordinary product replaced with the star product since,
as we can see from (\ref{ansatznBPS}), the world-volume
of  a non-BPS D(2p)-brane is noncommutative.
With using
\begin{equation}
G_s=g_s\sqrt{\frac{\det\lambda B}{\det g}} \ ,
\sqrt{\det{\lambda^{-1}\Theta}}=\lambda^{-p}
\sqrt{\det \Theta} \ ,
\end{equation}
we obtain  from  (\ref{actionfinal}) the action for
non-BPS D(2p)-brane 
\begin{eqnarray}\label{actionfinal2}
S=-\frac{C_{2p}}{G_s}
\int dt d^{2p}x
\sqrt{-\det\left(\lambda
(\hat{F}_{kl}-B_{kl})+
P[G_{kl}+G_{ka}(X^{ab}-\delta^{ab})G_{bl}]\right)}
\times \nonumber \\
\times \sqrt{\det(Q^a_b)}
\left(V(T)-\sum_{n=1}f_n(T)\lambda^n\left(
G^{ab}D_aTD_bT-g_{ij}
[\Phi^i,T][\Phi^j,T]\right)^n\right)
 \ , \nonumber \\
\end{eqnarray}
where we have used
\begin{equation}
\frac{C_0}{(2\pi \lambda)^p}=C_{2p} \ .
\end{equation}
We have seen that $N$ non-BPS D-branes
in the limit $N\rightarrow \infty$ have solution corresponding
to non-BPS D-branes of higher dimension. This
solution is in some sense dual to the tachyon condensation
on the world-volume of space-time filling branes with
non-commutative world-volume. In fact, there is a closed
relation between non-BPS D0-branes and action for
non-BPS D-brane written in operator formalism
\cite{StromingerA,Harvey, SenTP}. In this section we
have demonstrated this relation more explicitly.
The generalisation to the case of non-abelian non-BPS 
D(2p)-brane is straightforward \cite{Seiberg}.
\section{Conclusion}

In this short note we have presented some results
considering tachyon condensation in the system of $N$ 
non-BPS D0-branes. We have seen in the section (\ref{second})
that there is a solution with zero mass 
which  we have interpreted as a tensionless D0-brane
\cite{Harvey,Gopakumar}.
This problem is
similar to the problem of tensionless circular D8-brane in
\cite{WittenFT}. However, there is a puzzle.
 If this was genuine
light state in Type II string theory they should have been known
already from other studies.  At present we do not know how resolve
this puzzle. Resolution of this problem has been suggested in
\cite{Harvey2} in the framework of noncommutative
geometry which was based on the extra $Z_2$ discrete
gauge  symmetry
of the action for non-BPS D-brane. We hope to return to
this important question in the future.

In section (\ref{fourth}) we have proposed an ansatz
which leads to the emergence of BPS D-branes from non-BPS
D0-branes. Unfortunately, we were not able to calculate
the tension of resolution object directly, since we have used
the linear approximation only. However, from the form of
the energy of this configuration which scales as a energy
of BPS D-brane and also from the charge of resulting 
configuration we can claim the these solutions really 
describe BPS D-branes in Type IIB theory since non-BPS D0-branes
are present in Type IIB theory. It would be nice to go 
beyond linear approximation which seams to be 
impossible at present since we do not know the exact form
of the action. On the other hand, it was suggested in
\cite{SenTP} that it is possible that higher order terms in
the action are only gauge artefacts. It would be nice to
have some exact proof this intriguing conjecture.

We have also seen that from the action for non-BPS
D0-branes we can obtain action for higher dimensional
non-BPS D(2p)-branes in the very elegant way used
in the construction of higher dimensional branes in
Matrix theory and in Type IIB theory. We have seen
that this analysis is valid for the whole effective action
without restriction to the linear approximation. 
The same analysis could be possible with the Wess-Zumino
term for a non-BPS D0-branes which could lead to
the Wess-Zumino term for noncommutative D-branes
presented recently in the beautiful paper \cite{Mukhi}.
We hope to return to this question in the  future.

\newpage

\end{document}